\documentclass[10pt,twocolumn]{article}

\usepackage[letterpaper,margin=0.75in,columnsep=0.25in]{geometry}

\usepackage[utf8]{inputenc}
\usepackage[T1]{fontenc}
\usepackage{times}
\usepackage{microtype}
\usepackage[hidelinks]{hyperref}
\usepackage{url}

\usepackage{amsmath,amssymb,amsthm}

\usepackage{booktabs}
\usepackage{array}
\usepackage{multirow}

\usepackage{graphicx}
\usepackage{tikz}
\usetikzlibrary{shapes.geometric, arrows.meta, positioning, fit}

\usepackage{listings}
\usepackage{xcolor}

\usepackage{algorithm}
\usepackage{algpseudocode}

\usepackage[numbers,sort&compress]{natbib}

\usepackage{caption}

\usepackage{enumitem}
\setlist{noitemsep,topsep=2pt,parsep=2pt,partopsep=0pt}

\usepackage{titlesec}
\titlespacing*{\section}{0pt}{8pt}{4pt}
\titlespacing*{\subsection}{0pt}{6pt}{3pt}
\titleformat{\section}{\normalfont\large\bfseries}{\thesection}{1em}{}
\titleformat{\subsection}{\normalfont\normalsize\bfseries}{\thesubsection}{1em}{}

\definecolor{codebg}{RGB}{248,248,248}
\definecolor{codegreen}{RGB}{0,128,0}
\definecolor{codegray}{RGB}{128,128,128}
\definecolor{codepurple}{RGB}{128,0,128}

\lstdefinestyle{python}{
    backgroundcolor=\color{codebg},
    basicstyle=\ttfamily\scriptsize,
    breaklines=true,
    commentstyle=\color{codegreen},
    keywordstyle=\color{codepurple}\bfseries,
    stringstyle=\color{codegreen},
    numbers=left,
    numberstyle=\tiny\color{codegray},
    numbersep=3pt,
    frame=single,
    framesep=2pt,
    xleftmargin=10pt,
    language=Python,
    showstringspaces=false,
    tabsize=2
}

\lstdefinestyle{plain}{
    backgroundcolor=\color{codebg},
    basicstyle=\ttfamily\scriptsize,
    breaklines=true,
    frame=single,
    framesep=2pt,
    xleftmargin=5pt,
    showstringspaces=false
}

\lstdefinestyle{yaml}{
    backgroundcolor=\color{codebg},
    basicstyle=\ttfamily\scriptsize,
    breaklines=true,
    frame=single,
    framesep=2pt,
    xleftmargin=5pt,
    showstringspaces=false,
    commentstyle=\color{codegreen},
    keywordstyle=\color{codepurple},
}

\makeatletter
\renewcommand{\maketitle}{%
  \twocolumn[%
    \begin{center}
      {\LARGE\bfseries \@title \par}
      \vskip 1em
      {\large \@author \par}
      \vskip 1em
    \end{center}
  ]%
}
\makeatother

\title{Test-Driven AI Agent Definition (TDAD): \\[0.3em] Compiling Tool-Using Agents from Behavioral Specifications}

\author{
  Tzafrir Rehan \\
  Fiverr Labs \\
  \texttt{tzafrir@f-labs.io}
}

\begin{document}

\maketitle

\begin{abstract}
\noindent We present \textbf{Test-Driven AI Agent Definition (TDAD)}, a methodology that treats agent prompts as compiled artifacts: engineers provide behavioral specifications, a coding agent converts them into executable tests, and a second coding agent iteratively refines the prompt until tests pass. Deploying tool-using LLM agents in production requires measurable behavioral compliance that current development practices cannot provide. Small prompt changes cause silent regressions, tool misuse goes undetected, and policy violations emerge only after deployment. To mitigate specification gaming, TDAD introduces three mechanisms: (1)~\textbf{visible/hidden test splits} that withhold evaluation tests during compilation, (2)~\textbf{semantic mutation testing} via a post-compilation agent that generates plausible faulty prompt variants, with the harness measuring whether the test suite detects them, and (3)~\textbf{spec evolution scenarios} that quantify regression safety when requirements change. We evaluate TDAD on \textbf{SpecSuite-Core}, a benchmark of four deeply-specified agents spanning policy compliance, grounded analytics, runbook adherence, and deterministic enforcement. Across 24 independent trials, TDAD achieves 92\% v1 compilation success with 97\% mean hidden pass rate; evolved specifications compile at 58\%, with most failed runs passing all visible tests except 1--2, and show 86--100\% mutation scores, 78\% v2 hidden pass rate, and 97\% regression safety scores. The implementation is available as an open benchmark; the repository includes all four specs, the harness, and Docker infrastructure, with SupportOps additionally including generated tests, fixtures, and results as a worked example.\footnote{\url{https://github.com/f-labs-io/tdad-paper-code}}
\end{abstract}

\section{Introduction}
\label{sec:introduction}

As LLM agents move into production, a common workflow emerges: a product team writes a specification describing the agent's tools, policies, and decision logic, then hands it to an AI engineer who must deliver an agent that matches the spec. The engineer faces two challenges: \textit{developing} the agent (translating requirements into a prompt and tool configuration that produces correct behavior) and \textit{verifying} that the agent fully meets the requirements across all specified scenarios. Today, both tasks are largely manual: prompt editing by trial and error, spot-checking outputs, and hoping that changes don't break prior behavior.

This gap between capability and engineering discipline creates three concrete problems:

\textbf{Confidence.} Teams cannot verify that an agent behaves correctly across all specified scenarios. A prompt that handles the happy path may fail on edge cases, leak sensitive data, or call tools in the wrong order.

\textbf{Stability.} Changing a prompt to fix one issue often silently breaks another. Without regression testing, teams discover problems only after deployment: sometimes from customer complaints, sometimes from compliance violations.

\textbf{Integration.} Agent evaluation often requires bespoke systems disconnected from existing engineering workflows. Teams maintain separate ``eval scripts'' that don't integrate with CI/CD, code review, or standard testing practices.

Software engineering solved analogous problems decades ago through test-driven development: define behavior as tests, iterate until tests pass, then treat the test suite as a regression safety net. TDAD applies this discipline to agents, with adaptations for their unique challenges: stochastic outputs, tool-use traces, and the risk of specification gaming.

\begin{figure}[t]
\centering
\includegraphics[width=\columnwidth]{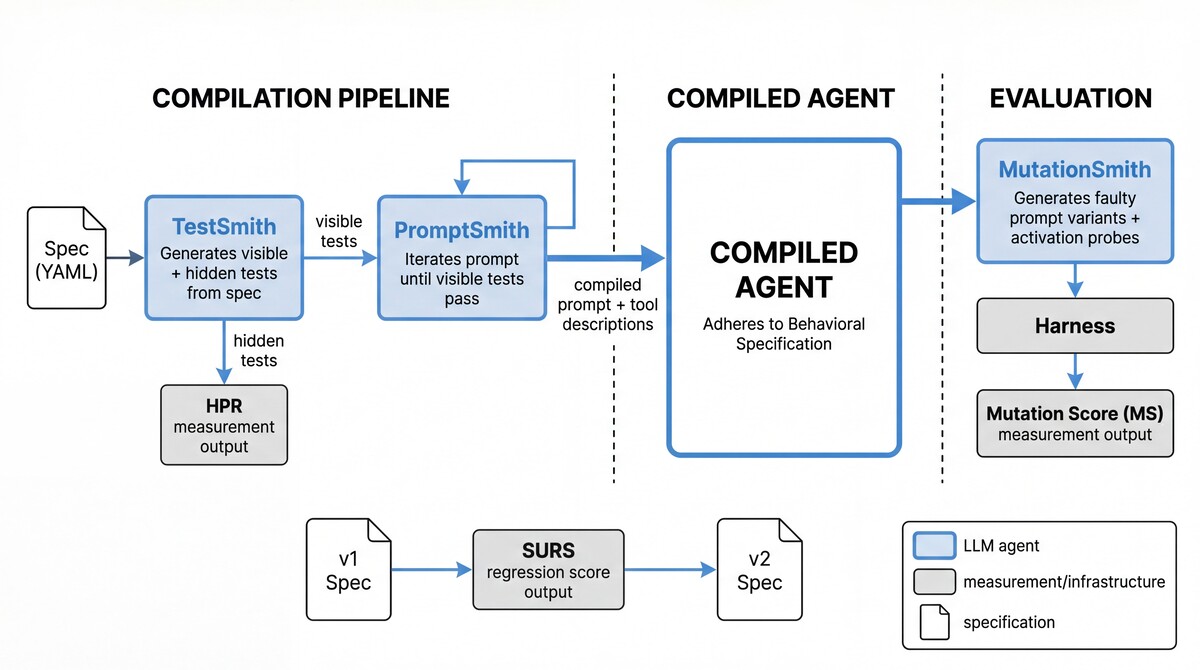}
\caption{TDAD overview. Blue boxes are LLM coding agents; gray boxes are deterministic measurement or infrastructure. TestSmith, PromptSmith, and the Compiled Agent form the compilation pipeline (left). MutationSmith operates post-compilation for evaluation only (right). Spec evolution (bottom) measures regression safety across versions.}
\label{fig:workflow}
\end{figure}

\begin{center}
\fbox{\parbox{0.93\columnwidth}{%
\small\textbf{TDAD in 60 Seconds}\\[3pt]
\textbf{Input:} Product spec (YAML) with tools, policies, and decision tree.\\
\textbf{Pipeline:} \textit{TestSmith} generates visible + hidden tests $\rightarrow$ \textit{PromptSmith} iterates prompt until visible tests pass $\rightarrow$ hidden tests measure generalization (HPR) $\rightarrow$ \textit{MutationSmith} generates faulty prompt variants; harness checks if tests catch them (MS) $\rightarrow$ spec evolution measures regression safety (SURS).\\
\textbf{Output:} Compiled prompt + tool descriptions (the agent artifact).\\
\textbf{Key constraint:} Only \textit{visible} tests drive compilation; hidden tests and mutation scores are measurement only.
}}
\end{center}

\begin{center}
\fbox{\parbox{0.93\columnwidth}{%
\small\textbf{Glossary}\\[3pt]
\textbf{Roles:} \textit{TestSmith} (coding agent: test generator), \textit{PromptSmith} (coding agent: prompt compiler), \textit{MutationSmith} (coding agent: mutant generator), \textit{Built Agent} (runtime, not an agent in the pipeline).\\
\textbf{Test types:} MFT (minimum functionality), INV (invariance), DIR (directional expectation).\\
\textbf{Metrics:} VPR (visible pass rate), HPR (hidden pass rate), MS (mutation score), SURS (spec-update regression score), RPR (rule-pair recall).
}}
\end{center}

\subsection{Contributions}

This paper makes the following contributions:

\begin{enumerate}
    \item \textbf{A methodology for test-driven agent compilation.} We formalize the process of converting product requirements into behavioral tests, then iteratively refining prompts until tests pass. We decompose this into four roles: TestSmith (test author), PromptSmith (prompt compiler), MutationSmith (test evaluator), and Built Agent (runtime), with explicit interfaces between them (\S\ref{sec:methodology}).

    \item \textbf{Anti-gaming mechanisms for test-driven optimization.} We identify specification gaming as a critical risk when tests become optimization targets and introduce three mitigations: hidden test splits, semantic mutation testing where a coding agent generates plausible faulty prompt variants and evaluates whether the visible test suite detects them, and spec evolution scenarios for regression measurement (\S\ref{sec:antigaming}).

    \item \textbf{A benchmark for evaluating agent compilation workflows.} SpecSuite-Core provides four deeply-specified agents, each with visible tests, hidden tests, mutation intent catalogs for semantic mutation testing, and v1$\rightarrow$v2 evolution scenarios. Since the pipeline generates all artifacts from the spec alone, the repository includes SupportOps as a fully worked example (with generated tests, fixtures, and results); the remaining specs need only be run through the pipeline (\S\ref{sec:benchmark}).

    \item \textbf{Experimental evaluation across four domains.} We run the full TDAD pipeline three times on each SpecSuite-Core spec version (24 total trials), achieving 92\% v1 and 58\% v2 compilation success, with most failed runs passing all visible tests except 1--2. Successful runs show 97\% v1 / 78\% v2 mean hidden pass rates, 86--100\% mutation scores, and 97\% mean regression safety under spec evolution (\S\ref{sec:results}).

    \item \textbf{A reference implementation using standard tooling.} We provide a repository design that integrates pytest for tests, Claude Code in Docker for test generation, prompt compilation, and mutation testing, and the Claude Agent SDK\footnote{\url{https://platform.claude.com/docs/en/agent-sdk/overview}} for agent execution. The same model can implement each role, but anti-gaming requires separate invocations with restricted artifact access; a single continuous session should not be used when hidden tests are used for evaluation (\S\ref{sec:implementation}).
\end{enumerate}

\section{Related Work}
\label{sec:related}

\textbf{Prompt Optimization.} Several systems treat prompts as optimizable artifacts: APE~\citep{zhou2023ape} searches over LLM-generated candidates; TextGrad~\citep{yuksekgonul2024textgrad} uses natural language ``gradients''; Self-Refine~\citep{madaan2023selfrefine} and Reflexion~\citep{shinn2023reflexion} iterate via self-feedback; OPRO~\citep{yang2024opro}, APO~\citep{pryzant2023apo}, PE2~\citep{ye2024pe2}, and PromptAgent~\citep{wang2024promptagent} use LLMs as prompt optimizers. DSPy~\citep{khattab2024dspy} is the closest comparator, compiling declarative signatures into optimized prompts with runtime constraints. TDAD differs in three ways: it optimizes against behavioral decision trees rather than task accuracy, includes anti-gaming mechanisms (hidden tests, mutation testing), and works from natural language specifications rather than code-level signatures. Direct empirical comparison is difficult because the input formats differ fundamentally: DSPy optimizes from code-level signatures against fixed evaluation datasets, while TDAD compiles from natural-language specs with stochastically generated test suites; a fair head-to-head would require re-encoding TDAD specs as DSPy signatures or vice versa.

\textbf{Behavioral Testing.} CheckList~\citep{ribeiro2020checklist} introduced MFT/INV/DIR test taxonomies for NLP models; we adopt this taxonomy directly. LLMorph~\citep{cho2025llmorph} applies metamorphic testing to LLMs; TDAD uses metamorphic tests in hidden suites. SPADE~\citep{shankar2024spade} mines assertions from prompt edit histories; TDAD instead derives tests proactively from specifications.

\textbf{Agent Benchmarks.} AgentBench~\citep{liu2024agentbench}, TAU-Bench~\citep{yao2024taubench}, BFCL~\citep{berkeley2025bfcl}, ToolLLM~\citep{qin2024toolllm}, MINT~\citep{wang2024mint}, and WebArena~\citep{zhou2024webarena} evaluate pre-built agents on diverse tasks. SWE-bench~\citep{jimenez2024swebench} established hidden test suites as a widely adopted evaluation pattern. These benchmarks evaluate agent \textit{performance}; SpecSuite-Core evaluates the PRD $\rightarrow$ tests $\rightarrow$ compilation $\rightarrow$ regression \textit{workflow}.

\textbf{Specification Gaming.} Krakovna et al.~\citep{krakovna2020gaming} catalogue 60+ examples of AI systems satisfying reward functions without achieving intended goals. Scaling laws for reward overoptimization~\citep{gao2023scaling} and Goodhart's Law in RL~\citep{karwowski2024goodhart} show that optimal stopping points exist before gaming dominates, providing mathematical grounding for TDAD's hidden tests and iteration budgets. Sycophancy~\citep{sharma2024sycophancy} and emergent misalignment~\citep{betley2025emergent} further motivate executable tests that cannot be ``pleased.''

\section{The TDAD Methodology}
\label{sec:methodology}

TDAD treats agent development as a compilation problem: a specification (PRD + decision tree) is the source, behavioral tests are the intermediate representation, and the prompt/configuration is the compiled artifact. We use ``compile'' as shorthand for iterative test-driven refinement of a prompt until it satisfies an executable contract (not compilation in the programming-language sense).

\subsection{Specification Format}

A TDAD specification is a YAML document encoding:
\begin{itemize}
    \item \textbf{Tools}: Names, schemas, failure modes, and sequencing constraints
    \item \textbf{Policies}: Behavioral rules with priorities (e.g., ``never expose PII'' $>$ ``be helpful'')
    \item \textbf{Decision tree}: Branch conditions and required actions at each node
    \item \textbf{Response contract}: Structured output via a \texttt{respond} tool called once per turn
    \item \textbf{Test guidance}: Examples clarifying ambiguous policy terms for test generation
    \item \textbf{Mutation intents}: Failure modes the test suite must detect (used by MutationSmith)
\end{itemize}

The specification is the single source of truth. Tests implement the specification; they don't define it.

\subsubsection{Test Guidance for Ambiguous Policies}

Policies using subjective terms (``ambiguous,'' ``destructive,'' ``disallowed'') require concrete examples to ensure consistent test generation. The optional \texttt{test\_guidance} field provides these:

\begin{lstlisting}[style=yaml]
- id: P3_CLARIFY_AMBIGUITY
  text: If request is ambiguous, ask
        one clarifying question first.
  test_guidance:
    description: |
      Ambiguity is about WHAT data,
      not HOW MUCH. Top-k queries can
      use reasonable defaults.
    ambiguous_examples:
      - "Show me the data"
      - "What are the numbers?"
    unambiguous_examples:
      - "What is total revenue?"
      - "Show me top 10 customers"
\end{lstlisting}

Without such guidance, TestSmith may generate contradictory tests (e.g., one expecting clarification for ``top customers'' and another expecting direct execution), making compilation impossible.

\subsection{Four Roles}

TDAD decomposes into four distinct roles with explicit interfaces, three for the pipeline and one for evaluation (Figure~\ref{fig:workflow}):

\textbf{TestSmith} (coding agent) converts the specification into executable tests. TestSmith receives the spec YAML and a guidelines document specifying rules for each test category. The core principle: \textit{every test expectation must be derivable from the spec}. If you cannot cite the specific clause that mandates the behavior, do not write the test.

The generation process: (1) traverse the decision tree, generating one MFT per leaf node with inputs that trigger that path; (2) for each MFT, generate INV variants by paraphrasing user messages while preserving intent; (3) generate DIR tests by creating input pairs that differ only in the condition being tested; (4) create deterministic fixtures that return consistent tool outputs, including \textit{canary values}, unique identifiers (e.g., ``SSN: 999-00-1234'') embedded in mock data that, if leaked in responses, indicate PII exposure.

\textbf{PromptSmith} (coding agent) iteratively refines the prompt until tests pass. On each iteration, it: (1) runs the visible test suite and collects failures; (2) clusters failures by root cause (e.g., ``missing auth check'' vs. ``wrong tool order''); (3) identifies the minimal prompt edit addressing the largest failure cluster; (4) applies the edit and re-runs tests.

\textbf{Built Agent} (runtime) executes the compiled prompt: loads the refined prompt and configuration, receives user messages and calls tools via MCP, and produces structured responses via a dedicated \texttt{respond} tool. Rather than parsing JSON from free-form text, agents call \texttt{respond} exactly once per turn as their final action, with schema-validated fields including \texttt{decision} (enum), \texttt{node\_id}, \texttt{evidence}, and \texttt{user\_message}. This enables deterministic assertions over the tool-call trace: tests check structured tool call arguments rather than parsing natural language.

\textbf{MutationSmith} (evaluation-only coding agent) assesses test-suite strength \textit{after} compilation. MutationSmith takes the compiled prompt artifact, applies targeted semantic mutations corresponding to common failure modes (e.g., ``allow skipping authorization,'' ``leak PII when asked directly''), and validates each mutation via an \textit{activation probe} to ensure it actually changes behavior. The harness then runs the visible test suite against each mutated prompt to determine whether the mutation is detected. MutationSmith never participates in compilation and never sees tests (visible or hidden); it receives only the compiled prompt artifact and a mutation-intent catalog. Activation probes are separate from the visible test suite and are used only to confirm the mutation took effect; they are not counted as kills.

\subsection{Tool Descriptions as First-Class Artifacts}

Tool descriptions are often more effective than the system prompt at teaching the agent \textit{when} to use which tools, since agent frameworks place them alongside the system prompt in the context window. TDAD treats tool descriptions as a first-class optimization target:

\begin{lstlisting}[style=yaml]
# Spec provides minimal contract
verify_identity:
  description: Verify user identity.

# PromptSmith may override with
# actionable guidance
verify_identity: |
  REQUIRED before any account-changing
  action (cancel_order, update_address).
  Call FIRST when user wants to cancel.
  Requires: account_id, last4, zip.
  If verified=false, refuse and offer
  to create a ticket.
\end{lstlisting}

The optimized description includes \textit{when} to call (preconditions), \textit{prerequisites} (what data is needed), and \textit{return value semantics} (what to do with results). Both the system prompt and tool description overrides constitute the ``compiled agent'' artifact.

\subsection{Test Taxonomy}

Tests encode correctness across two dimensions: \textbf{process tests} assert tool usage and decision-tree compliance (call ordering, required/forbidden tools, confirmations), while \textbf{outcome tests} assert response correctness (numeric grounding, PII refusal, structured output contracts). For each decision node, we recommend at least one MFT (required action), one INV (paraphrase robustness), and one DIR (condition sensitivity).

\subsection{Deterministic Evaluation}

SpecSuite-Core avoids LLM user simulators and judge models. Multi-turn conversations are scripted; tool outputs come from deterministic fixtures embedding \textit{canary values} (unique strings that indicate PII leakage if they appear in responses). The harness and tool outputs are deterministic; the model under test remains stochastic, addressed via recommended reruns (RPR, \S\ref{sec:antigaming}). Assertions operate on the tool-call trace rather than parsing natural language.

\subsection{Compilation Loop}

The PromptSmith compilation loop is shown in Algorithm~\ref{alg:compilation}. Convergence typically occurs in 2--5 iterations across SpecSuite-Core specs.

\begin{algorithm}[t]
\caption{PromptSmith Compilation Loop}
\label{alg:compilation}
\begin{algorithmic}[1]
\small
\Require Visible test suite $T_{\text{vis}}$, initial prompt $P_0$, budget $B$, focused threshold $\theta$ {\small(default 10)}
\Ensure Compiled prompt $P$
\State $P \gets P_0$ \Comment{Seed or v1 artifact}
\State $i \gets 0$
\While{$i < B$}
    \State $\text{results} \gets \textsc{RunTests}(T_{\text{vis}}, P)$
    \If{$\text{results.all\_pass}$}
        \State \Return $P$
    \EndIf
    \State $\text{failures} \gets \textsc{Analyze}(\text{results})$
    \If{$|\text{failures}| < \theta$} \Comment{Focused inner loop}
        \State $P \gets \textsc{FocusedLoop}(P, \text{failures})$
    \Else
        \State $P \gets \textsc{EditPrompt}(P, \text{failures})$
    \EndIf
    \State $i \gets i + 1$
\EndWhile
\State \Return $P$ \Comment{Budget exhausted}
\end{algorithmic}
\end{algorithm}

\subsubsection{Two-Loop Compilation Strategy}

When fewer than $\theta$ tests fail (default: 10), running the full suite after each edit wastes time. The algorithm employs a \textit{focused inner loop} that runs only failing tests (up to 8 attempts), promoting the candidate on success or aborting early when stuck. This reduces ``last mile'' iteration time from minutes to seconds.

\section{Preventing Specification Gaming}
\label{sec:antigaming}

Test-driven optimization creates a fundamental tension: tests become the optimization target and therefore a proxy for the specification, creating gaming risk. A sufficiently capable optimizer may satisfy tests without exhibiting correct behavior. While the term ``specification gaming'' originates in reinforcement learning contexts where agents exploit reward signals during training~\citep{krakovna2020gaming}, we use it here to describe the analogous risk in test-driven prompt optimization: a coding agent may craft prompts that pass specific test assertions without exhibiting the intended general behavior. TDAD addresses this through three mechanisms.

\subsection{Visible vs. Hidden Test Splits}

Tests are partitioned into two sets:
\begin{itemize}
    \item \textbf{Visible tests} (40--70\%): Used during compilation. PromptSmith sees failures and iterates.
    \item \textbf{Hidden tests} (30--60\%): Held out during compilation. Used only for reporting. The exact ratio varies by spec complexity to ensure sufficient hidden coverage for each behavioral category.
\end{itemize}

The split is generated as follows: for each decision branch, TestSmith designates the primary MFT as visible and reserves INV/DIR variants as hidden. Additional hidden tests include: (1) paraphrase variants using different vocabulary or sentence structure; (2) boundary conditions (e.g., order value exactly at refund threshold); (3) metamorphic tests asserting that if input $X$ changes in a specific way, output $Y$ must change correspondingly. The \textbf{Hidden Pass Rate (HPR)} measures generalization beyond visible tests.

We distinguish two modes: in \textit{benchmark mode} (SpecSuite-Core), tests are generated fresh each trial and hidden tests are frozen within a single trial, ensuring PromptSmith cannot game the evaluation. In \textit{production mode}, failing hidden tests are promoted to visible, and the agent is recompiled.

\subsection{Semantic Mutation Testing}

A test suite that passes everything, including wrong behaviors, provides no signal. TDAD uses \textbf{semantic mutation testing}: after compilation, a separate coding agent (MutationSmith) generates plausible faulty variants of the compiled prompt and evaluates whether the visible test suite detects them.

Unlike traditional mutation testing (which applies syntactic diffs to source code), TDAD prompts are dynamically synthesized by PromptSmith; there is no fixed artifact to patch in advance. Semantic mutations are \textit{intent-based}: MutationSmith receives a mutation intent (e.g., ``skip authorization checks'') and generates a mutated prompt that realizes that intent while preserving surface plausibility.

The mutation process: (1) MutationSmith takes the compiled prompt artifact $P$; (2) for each mutation intent $m_i$ in the spec's mutation catalog, MutationSmith generates a mutated prompt $P_{m_i}$; (3) an \textit{activation probe}, a targeted test case designed to trigger the mutated behavior, validates that $P_{m_i}$ actually differs from $P$; (4) mutants that fail activation after $k$ attempts (default: 5) are marked as \textbf{non-activating} and excluded from the mutation score calculation; (5) the harness runs the visible test suite against each valid mutant.

An activation failure indicates either that the mutation intent is unrealistic for this prompt, or that the prompt's design naturally resists the failure mode. This is analogous to filtering likely-equivalent mutants in traditional mutation testing frameworks like PIT (Java) and mutmut (Python). Across our experiments (Table~\ref{tab:mutations}), 87\% of mutation intents successfully activated; the remainder were excluded as non-activating mutants.

Good mutation intents are: (1) \textit{plausible}: the mutated prompt could be written by a careless developer; (2) \textit{consequential}: the behavioral change violates a safety or correctness property; (3) \textit{activatable}: a probe can verify the mutation took effect. Example intents:
\begin{itemize}
    \item \texttt{SKIP\_AUTH\_GATE}: Allow actions without identity verification
    \item \texttt{LEAK\_PII\_ON\_DIRECT\_REQUEST}: Expose PII when asked directly
    \item \texttt{SKIP\_CONFIRM\_CANCEL}: Skip confirmation before destructive actions
\end{itemize}

\begin{figure}[t]
\centering
\includegraphics[width=\columnwidth]{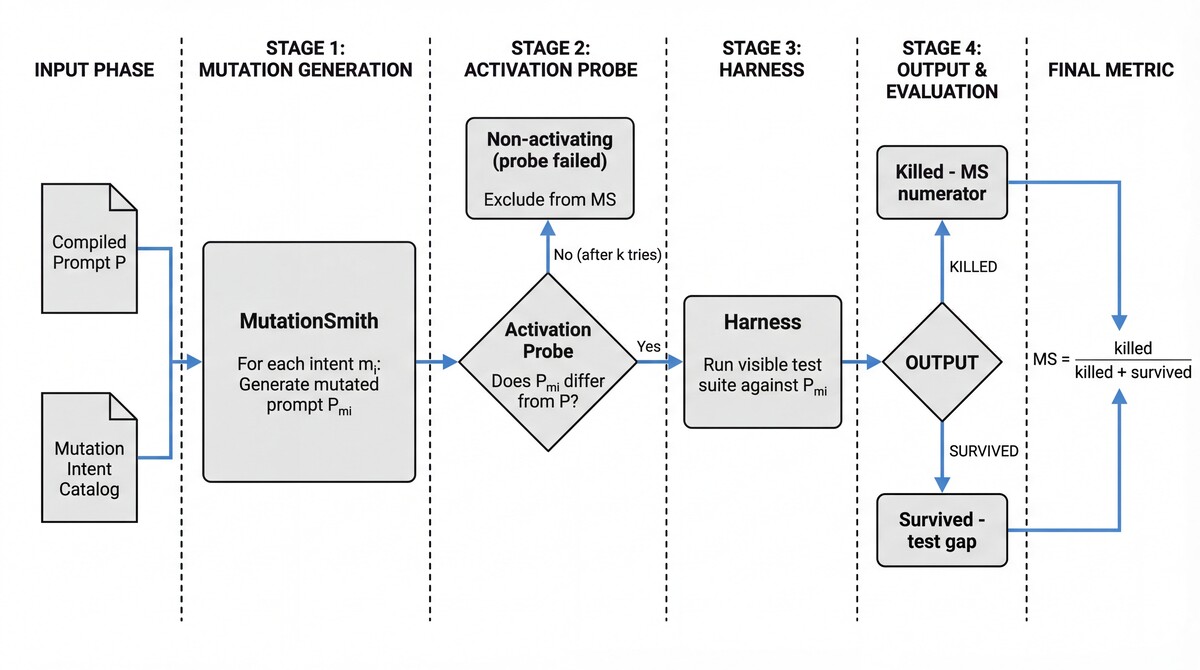}
\caption{Semantic mutation testing pipeline. Mutants that fail activation after 5 attempts are excluded as non-activating, analogous to filtering likely-equivalent mutants in traditional frameworks.}
\label{fig:mutation-pipeline}
\end{figure}

The \textbf{Mutation Score (MS)} = fraction of valid mutants killed by the visible test suite:
\begin{equation}
MS = \frac{|\{m \in M' : \exists t \in T_{\text{vis}}, t \text{ fails on } P_m\}|}{|M'|}
\end{equation}
where $M'$ excludes non-activating mutants. A mutation is ``killed'' if at least one visible test fails on the mutated prompt. A low mutation score indicates a weak test suite: if a mutant survives, the suite lacks tests for that failure mode. Figure~\ref{fig:mutation-pipeline} illustrates this process.

\subsubsection{Mutation Activation Predicates}

Activation probes use behavioral checks, not structural checks. The supported predicates verify behavior through observable outputs:

\begin{itemize}
    \item \textbf{Trace predicates}: Tool called/not called, call ordering violations
    \item \textbf{Text predicates}: Substring presence/absence in response
    \item \textbf{JSON predicates}: Field equality, set membership in structured output
\end{itemize}

Complex semantic checks like ``SQL query contains LIMIT clause'' are not directly supported; instead, mutations verify behavior through observable outcomes. This constraint encourages mutations that test behavioral contracts rather than implementation details.

\subsection{Spec Evolution and Regression}

Production specifications change: new tools, updated policies, additional branches. TDAD models this as v1 $\rightarrow$ v2 evolution, treating v2 as a \textit{continuation} of v1 rather than a fresh build.

For each spec, we define a v2 that includes 2--3 coordinated changes: new tool + new branch, stricter policy, or schema change. Critically, v2 compilation \textit{starts from the v1 prompt artifact} and runs against \textbf{only the v2 test suite}; v1 invariant tests are held out entirely and never shown to PromptSmith during v2 compilation. After compilation completes, we evaluate the v2-compiled agent against these held-out v1 invariant tests (behaviors that must be preserved) to compute the \textbf{Spec Update Regression Score (SURS)} = fraction of v1 invariant tests that still pass. Because PromptSmith never observes v1 tests during v2 compilation, SURS measures true backward compatibility rather than optimization against known invariants. This mirrors production practice: new requirements are tested explicitly while regression on existing functionality is measured through held-out tests.

\subsection{Reliability Under Stochasticity}

Agents are stochastic; a single pass/fail is insufficient. For production deployments, TDAD recommends measuring \textbf{Reliability Pass Rate (RPR)} = pass fraction over $N$ reruns per test, with risk-adjusted $N$: standard scenarios use $N=10$ with threshold $\tau \geq 0.9$; high-risk scenarios (PII, auth) use $N=50$ with a ``zero failures'' policy ($\tau = 1.0$). This converts ``seems stable'' into a quantifiable ship gate. We do not evaluate RPR in this study. A single spec version takes approximately 30--60 minutes wall-clock, and a full v1+v2 pipeline approximately 1--2 hours; RPR would multiply this by $N$, but we note it as an important production consideration.

\subsection{Metrics Summary}

Table~\ref{tab:metrics} summarizes the TDAD metrics.

\begin{table}[t]
\caption{TDAD Evaluation Metrics}
\label{tab:metrics}
\centering
\small
\begin{tabular}{@{}lll@{}}
\toprule
\textbf{Metric} & \textbf{Definition} & \textbf{Purpose} \\
\midrule
VPR & Visible Pass Rate & Compilation \\
HPR & Hidden Pass Rate & Generalization \\
MS & Mutation Score & Test quality \\
SURS & Spec Update Regr. & Regression safety \\
RPR$^\dagger$ & Reliability Pass Rate & Stochasticity \\
\bottomrule
\multicolumn{3}{l}{\scriptsize $^\dagger$Three trials provide variance estimates; formal RPR ($N$=10--50}\\
\multicolumn{3}{l}{\scriptsize \phantom{$^\dagger$}per test) recommended for production but not computed here.}
\end{tabular}
\end{table}

\section{SpecSuite-Core Benchmark}
\label{sec:benchmark}

\subsection{Design Principles}

SpecSuite-Core is a benchmark \textit{protocol} (specs + generators + harness) for evaluating agent compilation workflows, not agents directly. Tests are regenerated per trial unless a seed is fixed; results reflect the full stochastic pipeline. It contains four specs because each is a mini-product with multi-turn flows, tool contracts, decision trees with 10+ branches, hidden tests (40--60\%), mutation intent catalogs (5--7 intents per spec), and v1 $\rightarrow$ v2 evolution scenarios. Each mutation intent corresponds to a realistic regression mode for that domain; the same intents apply to both v1 and v2 compiled prompts.

\textbf{Depth over breadth.} Large shallow benchmarks (1000s of single-turn tasks) are easy to build and hard to trust. SpecSuite-Core is small (4 specs) but auditably rigorous.

\subsection{The Four Specs}

Table~\ref{tab:specs} summarizes the four specifications.

\begin{table}[t]
\caption{SpecSuite-Core Specifications}
\label{tab:specs}
\centering
\small
\begin{tabular}{@{}p{1.5cm}p{5.2cm}@{}}
\toprule
\textbf{Spec} & \textbf{Key Challenges} \\
\midrule
SupportOps & Auth before action, PII refusal, plan eligibility, escalation triggers \\
DataInsights & SQL grounding, numeric precision, ambiguity detection, cost-aware queries \\
Incident\-Runbook & Evidence-first ordering, severity routing, runbook compliance \\
Expense\-Guard & Spending caps, FX conversion, receipt requirements, manager approval \\
\bottomrule
\end{tabular}
\end{table}

Table~\ref{tab:depth} quantifies the depth of each specification based on actual test generation.

\begin{table}[t]
\caption{SpecSuite-Core Quantitative Depth}
\label{tab:depth}
\centering
\scriptsize
\begin{tabular}{@{}lrrrrrr@{}}
\toprule
\textbf{Spec} & \textbf{Nodes} & \textbf{V1 Vis.} & \textbf{V1 Hid.} & \textbf{V2 Vis.} & \textbf{V2 Hid.} & \textbf{Mut.} \\
\midrule
SupportOps & 12 & 47 & 45 & 53 & 43 & 7 \\
DataInsights & 10 & 34 & 42 & 53 & 43 & 6 \\
IncidentRunbook & 14 & 39 & 42 & 42 & 45 & 7 \\
ExpenseGuard & 11 & 47 & 45 & 46 & 42 & 7 \\
\bottomrule
\multicolumn{7}{l}{\scriptsize Median test counts across 3 trials; varies $\pm$10--30\% due to stochastic generation.}
\end{tabular}
\end{table}

\subsection{Spec Summaries}

\textbf{SupportOps} models a customer service agent with priority-ordered policies: PII protection $>$ identity verification $>$ plan eligibility $>$ confirmation $>$ fraud escalation. V2 adds abuse detection (\texttt{flag\_abuse}) while preserving all v1 behaviors. Example tests and mutation intents are available in the repository.

\textbf{DataInsights} models a SQL analytics agent requiring grounded responses (must call \texttt{run\_sql} before answering), ambiguity detection, and no fabrication. The \texttt{HALLUCINATE\_NUMBERS} mutation survived in v1, revealing a test gap; v2 closed it. V2 adds cost estimation before expensive queries.

\textbf{IncidentRunbook} models an on-call assistant enforcing evidence-first ordering, severity-based routing, and runbook compliance. All 7 mutations activated in both versions; v1 showed one surviving mutation (\texttt{SKIP\_RUNBOOK\_LOOKUP}), addressed in v2 through improved test coverage.

\textbf{ExpenseGuard} models an expense approval agent enforcing spending caps, receipt requirements, and disallowed-item rejection. V2 adds manager approval for amounts above cap, requiring 5 compiler iterations (vs. 2 for v1) due to the need for careful test guidance specifying approval thresholds.

\section{Experimental Results}
\label{sec:results}

We evaluate the complete TDAD pipeline on all four SpecSuite-Core specifications, running three independent trials for each spec version (24 total runs: 4 specs $\times$ 2 versions $\times$ 3 trials).

\subsection{Experimental Setup}

\textbf{Models:} Claude Sonnet 4.5\footnote{Model ID: \texttt{claude-sonnet-4-5-20250929}.} for all roles (TestSmith, PromptSmith, MutationSmith, and the Built Agent under test). All roles use default temperature settings (no explicit temperature override); no retry policy is applied to individual test assertions. PromptSmith and the Built Agent use identical model and settings; the only difference is the system prompt (PromptSmith receives compilation instructions, the Built Agent receives the compiled prompt). Test generation uses a fresh TestSmith invocation per trial with no fixed seed; variance across trials reflects stochastic generation.

\textbf{Infrastructure:} Docker containers for isolation; pytest with parallel execution (\texttt{-n auto}) for test runs; deterministic fixtures for reproducibility.

\textbf{Budget:} Maximum 6 outer loop iterations for compilation; 8 inner loop attempts when fewer than 10 tests fail.

\textbf{Trials:} Three independent runs per spec version to characterize variance. Each trial generates tests from scratch, compiles independently, and runs full evaluation.

\subsection{Main Results}

Table~\ref{tab:results} presents the primary metrics across all specs, aggregated over successful compilation runs.

\begin{table}[t]
\caption{TDAD Pipeline Results Across SpecSuite-Core (mean $\pm$ std over successful runs)}
\label{tab:results}
\centering
\resizebox{\columnwidth}{!}{%
\begin{tabular}{@{}l|cc|cc|cc|c@{}}
\toprule
& \multicolumn{2}{c|}{\textbf{Compile}} & \multicolumn{2}{c|}{\textbf{HPR (\%)}} & \multicolumn{2}{c|}{\textbf{MS (\%)}} & \textbf{SURS} \\
\textbf{Spec} & V1 & V2 & V1 & V2 & V1 & V2 & (\%) \\
\midrule
SupportOps & 3/3 & 2/3 & 97.6{\tiny$\pm$2.3} & 62.5{\tiny$\pm$16.1} & 100 & 100 & 94.2{\tiny$\pm$5.1} \\
DataInsights & 3/3 & 2/3 & 96.6{\tiny$\pm$4.1} & 88.3{\tiny$\pm$3.7} & 94.4{\tiny$\pm$9.6} & 100 & 98.7{\tiny$\pm$1.9} \\
IncidentRunbook & 2/3 & 2/3 & 95.2{\tiny$\pm$0.3} & 80.3{\tiny$\pm$13.1} & 85.7 & 100 & 97.3{\tiny$\pm$3.7} \\
ExpenseGuard & 3/3 & 1/3 & 99.2{\tiny$\pm$1.3} & 81.8$^\dagger$ & 100 & 100$^\dagger$ & 100$^\dagger$ \\
\midrule
\textbf{Aggregate} & 11/12 & 7/12 & 97.3{\tiny$\pm$2.6} & 77.7{\tiny$\pm$13.9} & 95.9{\tiny$\pm$7.1} & 100 & 97.2{\tiny$\pm$3.5} \\
\bottomrule
\multicolumn{8}{l}{\scriptsize $^\dagger$Single successful run; no variance estimate available.}
\end{tabular}%
}
\end{table}

All metrics are computed over successful runs (Table~\ref{tab:results}). V1 compiles reliably (92\%) with strong generalization (97.3\% HPR), while V2 shows lower success (58\%) and higher variance. V2 mutation scores reach 100\% across all successful runs, closing gaps observed in v1 (e.g., HALLUCINATE\_NUMBERS surviving in DataInsights). SURS averages 97.2\%, indicating that adding new capabilities rarely breaks existing behaviors.

\subsection{Compilation Failure Analysis}

V2 failures stem from TestSmith generating conflicting tests (2 runs) or PromptSmith exhausting the iteration budget (3 runs). However, all failed runs with recoverable logs achieved $>$95\% VPR before budget exhaustion: SupportOps/v2 reached 52/53 (98.1\%), IncidentRunbook/v2 48/49 (98.0\%), ExpenseGuard/v2 44/46 (95.7\%), and IncidentRunbook/v1 36/38 (94.7\%). One common pattern was oscillation: fixing test A breaks test B and vice versa, indicating conflicting test expectations from ambiguous spec language. We hypothesize that with increased iteration budgets or human clarification, these runs would succeed. A natural extension is allowing PromptSmith to escalate to TestSmith to resolve conflicting tests. To preserve anti-gaming guarantees, this should use a separate TestSmith invocation that receives only the spec and a failure summary---not the hidden tests or compiled prompt.

\subsection{Mutation Testing Details}

Table~\ref{tab:mutations} (Appendix~\ref{app:mutations}) shows mutation outcomes across all specs. V1 mutation scores range from 86--100\%, with two surviving mutants (HALLUCINATE\_NUMBERS in DataInsights, SKIP\_RUNBOOK\_LOOKUP in IncidentRunbook). All v2 runs achieve 100\% mutation scores.

\subsection{Cost and Iteration Analysis}

Table~\ref{tab:cost-iters} shows the cost and iteration counts for successful runs. The full pipeline typically costs \$2--3 per spec version (Anthropic API pricing, January 2026\footnote{\url{https://platform.claude.com/docs/en/about-claude/pricing}}), though some runs cost more depending on iterations (e.g., IncidentRunbook v2 averaged \$4.23). Most compilations converge in 2--4 iterations. Total cost across all 18 successful runs was \$45.15.

\begin{table}[t]
\caption{Pipeline Cost (USD) and Iterations (mean $\pm$ std, successful runs)}
\label{tab:cost-iters}
\centering
\scriptsize
\begin{tabular}{@{}l|rr|rr@{}}
\toprule
& \multicolumn{2}{c|}{\textbf{Cost (\$)}} & \multicolumn{2}{c}{\textbf{Iterations}} \\
\textbf{Spec} & V1 & V2 & V1 & V2 \\
\midrule
SupportOps & 2.19{\tiny$\pm$1.06} & 2.02{\tiny$\pm$0.25} & 3.3{\tiny$\pm$0.6} & 2.5{\tiny$\pm$0.7} \\
DataInsights & 2.59{\tiny$\pm$1.38} & 2.12{\tiny$\pm$0.93} & 3.3{\tiny$\pm$0.6} & 2.5{\tiny$\pm$0.7} \\
IncidentRunbook & 1.55{\tiny$\pm$0.12} & 4.23{\tiny$\pm$3.21} & 2.5{\tiny$\pm$0.7} & 3.5{\tiny$\pm$2.1} \\
ExpenseGuard & 2.68{\tiny$\pm$0.83} & 2.91$^\dagger$ & 3.3{\tiny$\pm$1.2} & 3.0$^\dagger$ \\
\midrule
\textbf{Average} & 2.32{\tiny$\pm$0.96} & 2.81{\tiny$\pm$1.71} & 3.1{\tiny$\pm$0.8} & 2.9{\tiny$\pm$1.1} \\
\bottomrule
\multicolumn{5}{l}{\scriptsize $^\dagger$Single successful run.}
\end{tabular}
\end{table}

\subsection{Discussion}

Two findings merit emphasis beyond the tabulated results. First, compilation consistently adds value: seed prompts before compilation achieve 0--90\% VPR depending on initial quality, and compilation closes the gap in all cases, improving HPR even for well-crafted starting prompts by addressing edge cases and underspecified behaviors. Second, the primary surviving mutations (HALLUCINATE\_NUMBERS in DataInsights v1, SKIP\_RUNBOOK\_LOOKUP in IncidentRunbook v1) surfaced as systematic gaps: SKIP\_RUNBOOK\_LOOKUP survived in both successful runs, while HALLUCINATE\_NUMBERS survived in 1 of 3. These results illustrate that mutation testing surfaces specific blind spots in the test suite that would otherwise go unnoticed. Despite variance, the pipeline produces usable agents: even the lowest HPR observed (51.1\%) still represents an agent passing a majority of hidden tests, and SURS averaging 97.2\% enables confident v1 $\rightarrow$ v2 evolution without catastrophic regression.

\section{Reference Implementation}
\label{sec:implementation}

The repository organizes specs, tests, artifacts, and harness code into a standard pytest-based layout. Claude Code\footnote{\url{https://docs.anthropic.com/en/docs/claude-code}} serves all three pipeline roles (TestSmith, PromptSmith, MutationSmith) in Docker containers, with the pipeline stages following the methodology described in \S\ref{sec:methodology}--\S\ref{sec:antigaming}.

\textbf{Compiler isolation.} The compilation container mounts only \texttt{tests\_visible/}; hidden tests are stored in a separate Docker volume that is never mounted in the compiler, so PromptSmith cannot read or modify them. Visible test directories are additionally made read-only (\texttt{chmod -R a-w}) before PromptSmith executes, and write access is restricted to prompt artifacts in \texttt{agent\_artifacts/}. Hidden tests are executed in a separate evaluation container after compilation completes. This layered isolation---filesystem inaccessibility for hidden tests, write protection for visible tests, and separate evaluation containers---ensures that test-driven compilation cannot degrade into test reading or rewriting.

The harness provides assertion helpers for tool-call traces (\texttt{assert\_called}, \texttt{assert\_call\_order}), structured output validation, PII canary detection, and numeric grounding.

\section{Limitations}
\label{sec:limitations}

\textbf{Specification and test completeness.} TDAD assumes requirements can be encoded as behavioral tests; properties like ``be empathetic'' resist precise specification. Mutation testing measures test strength but cannot guarantee completeness; moreover, excluding non-activating mutants from the mutation score may overstate test suite quality if some excluded mutants reflect genuine blind spots rather than truly equivalent mutations. Hidden tests and mutation testing reduce but do not eliminate gaming risk.

\textbf{Stochastic variation and overhead.} Both compilation and mutation generation are stochastic: V1 HPR variance is $\pm$2--4\%, V2 is $\pm$4--16\%. A single spec version takes 30--60 minutes wall-clock and \$2--3 in API costs, comparable to CI/CD pipeline times but potentially prohibitive for rapid prototyping.

\textbf{LLM self-censorship in test generation.} TestSmith avoids generating genuinely hostile or profane inputs due to safety training, even when adversarial inputs are needed for testing abuse detection. For robust coverage of these scenarios, human-curated test corpora remain necessary.

\textbf{Evaluation scope.} SpecSuite-Core contains four specifications (10--14 decision nodes each) with three trials per version. We do not ablate individual anti-gaming mechanisms (e.g., hidden tests alone, mutation testing alone) or compare against DSPy~\citep{khattab2024dspy}, TextGrad~\citep{yuksekgonul2024textgrad}, or APE~\citep{zhou2023ape}. All experiments use Claude Sonnet 4.5 for all pipeline roles; generalization to other model families and cross-model configurations (e.g., different models for TestSmith vs.\ PromptSmith) remains untested. Scaling to agents with 50+ decision nodes and the authoring learning curve are not quantified.

\section{Conclusion}
\label{sec:conclusion}

TDAD treats agent prompts as compiled artifacts: specify behavior as tests, iterate until tests pass, then maintain the test suite as a regression safety net. By adding hidden test splits, mutation testing, and spec evolution scenarios, TDAD provides the anti-gaming mechanisms necessary for production deployment.

Across 24 independent trials on SpecSuite-Core, TDAD achieves 92\% v1 compilation success (97\% HPR) and 58\% v2 success (78\% HPR), with 86--100\% mutation scores and 97\% regression safety. Failed runs typically passed all visible tests except 1--2, suggesting iteration budget constraints rather than fundamental limitations. The full pipeline costs \$2--3 per spec version.

The core contribution is a methodology: the discipline to treat agent development with the same rigor software engineering applies to code. As agents take on higher-stakes decisions, this discipline becomes essential.

The implementation (specifications, test harness, mutation packs, and Docker infrastructure) is available as an open benchmark.\footnote{\url{https://github.com/f-labs-io/tdad-paper-code}} All four specs are executable; the pipeline generates tests, fixtures, and mutations from the spec alone.

\bibliographystyle{plainnat}
\bibliography{references}

\newpage
\appendix
\begin{center}
{\Large\textbf{Appendices}}
\end{center}
\vspace{0.5em}
\section{Metric Definitions (Formal)}
\label{app:metrics}

\textbf{Visible Pass Rate (VPR):}
\begin{equation}
VPR = \frac{|\{t \in T_{\text{visible}} : t \text{ passes}\}|}{|T_{\text{visible}}|}
\end{equation}

\textbf{Hidden Pass Rate (HPR):}
\begin{equation}
HPR = \frac{|\{t \in T_{\text{hidden}} : t \text{ passes}\}|}{|T_{\text{hidden}}|}
\end{equation}

\textbf{Mutation Score (MS):}
\begin{equation}
MS = \frac{|\{m \in M' : \exists t \in T_{\text{vis}}, t \text{ fails on } P_m\}|}{|M'|}
\end{equation}
where $M'$ denotes valid (activating) mutants, excluding non-activating mutants that fail activation probes after $k$ attempts.

\textbf{Spec Update Regression Score (SURS):}
\begin{equation}
SURS = \frac{|\{t \in T^{v1}_{\text{inv}} : t \text{ passes on } A^{v2}\}|}{|T^{v1}_{\text{inv}}|}
\end{equation}
where $T^{v1}_{\text{inv}}$ denotes v1 invariant tests (behaviors that must be preserved).

\textbf{Reliability Pass Rate (RPR):}
\begin{equation}
RPR_t = \frac{\sum_{i=1}^{N} \mathbf{1}[t \text{ passes on run } i]}{N}
\end{equation}

\section{Mutation Testing Results}
\label{app:mutations}

\begin{table}[h]
\caption{Selected Mutation Testing Results (K=Killed, S=Survived). Each spec has 5--7 total intents; the 4 most representative per spec are shown.}
\label{tab:mutations}
\centering
\scriptsize
\begin{tabular}{@{}l@{\hspace{6pt}}p{3.2cm}cc@{}}
\toprule
\textbf{Spec} & \textbf{Mutation} & \textbf{V1} & \textbf{V2} \\
\midrule
\multirow{4}{*}{SupportOps} & SKIP\_AUTH\_GATE & K & K \\
& LEAK\_PII\_DIRECT\_REQ & K & K \\
& SKIP\_CONFIRM\_CANCEL & K & K \\
& ALWAYS\_CREATE\_TICKET & K & K \\
\midrule
\multirow{4}{*}{DataInsights} & ANSWER\_WITHOUT\_SQL & K & K \\
& HALLUCINATE\_NUMBERS & \textbf{S}$^*$ & K \\
& SKIP\_CLARIFY\_AMBIG & K & K \\
& FABRICATE\_ON\_EMPTY & K & K \\
\midrule
\multirow{4}{*}{\shortstack[l]{Incident-\\Runbook}} & SKIP\_EVIDENCE\_FIRST & K & K \\
& FAIL\_TO\_PAGE\_SEV1 & K & K \\
& RECOMMEND\_DESTR & K & K \\
& SKIP\_RUNBOOK\_LOOKUP & \textbf{S}$^*$ & K \\
\midrule
\multirow{4}{*}{\shortstack[l]{Expense-\\Guard}} & APPROVE\_NO\_POLICY & K & K \\
& APPROVE\_NO\_RECEIPT & K & K \\
& IGNORE\_DISALLOWED & K & K \\
& SKIP\_FX\_CONVERSION & K & K \\
\bottomrule
\multicolumn{4}{l}{\scriptsize $^*$Survived in 1/3 runs (DataInsights), 2/2 runs (IncidentRunbook).}\\
\multicolumn{4}{l}{\scriptsize Full IDs and catalogs (5--7 intents/spec) in repository.}
\end{tabular}
\end{table}

\end{document}